\documentclass[12pt,aps,amsmath,latexsym,amsfonts]{JHEP3}

\usepackage{epsfig}
\usepackage{amsfonts,amssymb,amsmath}
\usepackage[hang]{subfigure}

\newcommand{\be}{\begin{equation}}
\newcommand{\ee}{\end{equation}}
\newcommand{\bea}{\begin{eqnarray}}
\newcommand{\eea}{\end{eqnarray}}

\def\Ab{{\bf A}}
\def\Bb{{\bf B}}
\def\ab{{\bf a}}
\def\bb{{\bf b}}

\def\r{{\bf r}}

\title{Kinks, extra dimensions, and gravitational waves }
\author{Eimear O'Callaghan\!
\thanks{Email: e.e.o'callaghan@durham.ac.uk}\ ,
Ruth Gregory\!
\thanks{Email: r.a.w.gregory@durham.ac.uk}\\
Institute for Particle Physics Phenomenology
and Centre for Particle Theory, \\
Durham University, South Road, Durham, DH1 3LE, UK}

\abstract{
We investigate in detail the gravitational wave signal from kinks
on cosmic (super)strings, including the kinematical effects from
the internal extra dimensions. We find that the signal is suppressed,
however, the effect is less significant that that for cusps.
Combined with the greater incidence
of kinks on (super)strings, it is likely that the kink signal offers
the better chance for detection of cosmic (super)strings.}

\preprint{IPPP-10/82\\ DCPT-10/164}

\begin{document}

\section{Introduction}

Cosmic strings, \cite{CS}, continue to play an important 
role in the diagnostic testing of high energy physics in the early universe. 
Initially, these strings were seen as relic topological defects of some 
Grand Unified phase transition in the early universe, \cite{ZelKib}, and it
was hoped that they would provide a causal alternative to inflation
for the generation of primordial perturbations. However, it was
rapidly realised that the perturbation spectrum predicted by
strings was in contradiction to the microwave background 
measurements, \cite{CMBrout}, and for a while, cosmic strings were relegated
to the ``also-rans'' of early universe cosmology. Recently however,
cosmic strings have come back to the fore as by-products of the
brane inflationary scenario in string theory \cite{braneinf,KKLMMT,BI,BCS,JST}.

The basic cosmological picture is that cosmic strings arise
during some phase transition in the early universe, \cite{ZelKib}.
By their nature, they are topologically stable, and can therefore
persist as a network of loops and long string. 
From the cosmological point of view, the internal structure of the 
very thin string is irrelevant, and it is modelled by 
a zero thickness line-like object whose motion is determined by the 
Nambu action, \cite{NG,FWC}:
\be
S = -\mu \int d^2 \sigma \; \sqrt{\gamma}
\ee
where $\mu$ is the mass per unit length of the string, and $\gamma$
the induced metric on the worldsheet.
Together with rules for intercommutation \cite{IC}, or how
crossing strings interact, this
gives the basic physics of how a network of cosmic strings will evolve.
Incorporating gravitational effects via a linearised approximation
indicates how fast energy is lost from the network, \cite{gwave}, and
putting all these pieces together gives the scaling
picture of the original cosmic string scenario \cite{NET}.
  
Cosmic (super)strings\footnote{We use the conventional
bracketed (super) to indicate
that these strings are neither genuine superstrings, nor old fashioned
cosmic strings living in four dimensions (4D), but are some
approximate line like object with some number of effective 
extra internal dimensions in which they can move.}, \cite{CSSrev},
are generally modelled in a similar fashion to cosmic strings, but
with one or two key differences arising because they are derived from
a higher dimensional theory, string theory, and have additional physics
arising as a result. As was realised early on, these (super)strings need
not intercommute when they apparently intersect in our 4D world, since
they can miss each other in the internal dimensions \cite{ICSC}, 
resulting in a denser network, \cite{AS2}. Another key
difference is that the (super)strings can move in the internal dimensions,
thus altering the effective kinematics. 

It is the effect of these extra dimensions on the kinematics,
and their consequences that we are interested in. The Nambu action
leads to a simple set of equations whose solutions are left and
right moving waves along the string, \cite{KT}. Mostly, the (super)string moves
in a nonsingular, albeit highly relativistic, fashion. However, 
there are two main exceptions: cusps and kinks\footnote{Certain cosmic 
(super)strings can also have junctions, \cite{ICSC,GWCSS,JCT}, however
we will not consider this additional feature in detail here.}. Cusps
are transitory events, where constructive interference between
the left and right moving modes causes a point of the (super)string to
instantaneously reach the speed of light. Kinks on the other hand
are relics of intercommutation, they are sharp `corners' on the
(super)string and represent a discontinuity in the wave velocity of 
either a left or right mover; because they are a feature of only one
wavefront, they persist and move along the (super)string, \cite{ACKinks}. 
Since both kinks and cusps represent
a certain level of singularity of the worldsheet swept out by
the (super)string, when including gravitational backreaction they
generate strong and distinct gravitational wave signals, 
\cite{gwave,FAN,GWL,Eloss,DV,DV2}.

Typically, gravitational wave emission from (super)string loops
is dominated by low frequency radiation, and contributes to the decay process
of loops and the general gravitational wave background. However,
in an important paper, Damour and Vilenkin (DV), \cite{DV}, showed 
that for cusps and kinks, the high frequency radiation was not
exponentially damped, but had a significant power law tail, and 
hence a significant burst of power. They showed how to compute
the gravitational waveform, and that the cusp/kink had a characteristic
beaming pattern in which radiation was narrowly focussed into a 
cone/fan associated with the wavevectors of the cusp/kink respectively.

In recent papers, \cite{CCGGZ}, we explored the impact of 
the extra internal dimensions for the (super)string motion. 
As first pointed out in \cite{AS}, these ``slow down'' 
the (super)string from our 4D perspective: this obviously
has an impact on the definition and likelihood of a cusp, 
which requires that the (super)string instantaneously moves at the speed
of light. In \cite{CCGGZ} we showed how these internal degrees of 
freedom alter the gravitational wave signal from cusps; the
effect proved to be significant: contrary to previous expectations, 
\cite{JST,DV2,BCS3}, the gravitational radiation signal is not 
enhanced by the lowering of the intercommutation probability 
in extra dimensions, but is in fact suppressed by 
several orders of magnitude, mainly due to a probabilistic effect
from the altered kinematics. We had initially focussed on cusps
as they were expected to give a much stronger signal than kinks, 
but clearly, with the cusp signal so suppressed, 
a re-visiting of the kink signal is in order.

The key features of the kink are its sharp nature and persistence.
As shown by Damour and Vilenkin (DV), \cite{DV}, (see also \cite{FAN})
a kink gravitational wave burst (GWB) looks like a ``fan'',
i.e.\ strongly localised in one angular direction, but spreading out
in the other. Thus, although the high frequency tail of the GWB has
a stronger fall-off than the cusp, it is radiated over a greater solid
angle and is therefore still relevant. Moreover, if, as argued in
\cite{BBHS}, kinks are prolific on cosmic (super)strings, it
is possible that the effect of proliferation can lift the kink
signal above that of the cusp.
Finally, since kinks only require a discontinuity in the 
wave velocity, and since cosmic (super)strings still intercommute, 
albeit with a reduced probability, the mechanism for kink
formation would appear to be robust, and thus some of the
probabilistic suppression that we saw in the case of cusps
may not be a factor in the case of kinks.

In this paper we calculate how the extra dimensions affect the kink 
signal. The impact of including the internal degrees of freedom is
to narrow the spread of the ``fan"; there is no corresponding
parameter space reduction of the kink probability for the reasons
mentioned above. Thus, the damping of the kink signal is considerably
less pronounced than that of the cusp, and our conclusion is that
it is the kink, and not the cusp, GWB's that are most likely to be
seen by LIGO or LISA. We also look carefully at the geometry of
the kink burst by analysing its magnitude, and show that it is a 
genuine fan, in that it has finite angular extent in both directions, 
although is less sharply cut-off in its extended direction. As a
by-product of our analysis, we correct the approximations
used thus far in the literature for the amplitude, \cite{BBHS,ED}.

\section{Cosmic string kinks in four dimensions}
\label{sec:3d}

We briefly review standard cosmic string kinematics from the 
formalism developed by Kibble and Turok, \cite{KT}, then 
the DV calculation of the kink GWB. Finally, we discuss the amplitude
and geometry of the kink GWB and give an analysis of the 
actual contribution to the GW signal from a generic kink.

Writing the spacetime coordinates of the string as
$X^\mu(\sigma^A)$, where $\sigma^A=\{ \tau, \sigma\}$ are intrinsic
coordinates on the worldsheet, with $\sigma \in [0,L]$, 
for closed loops of length $L$. Kibble and Turok chose a gauge
with conformally flat worldsheet metric, centre of mass 
spacetime coordinates, and worldsheet time corresponding to coordinate
time. Writing $X^\mu=(\tau, {\bf r}(\tau, \sigma))$, they found
the solutions to the equation of motion for the worldsheet take the
form of left and right moving waves, conventionally written in the form
\be
\r= \frac{1}{2}[\ab(\sigma_-)+\bb(\sigma_+)],
\ee
where $ \sigma_\pm = \tau\pm \sigma$ are lightcone coordinates, and
the gauge conditions constrain $\ab'$ and $\bb'$ to lie on a unit
sphere, commonly dubbed the ``Kibble-Turok'' sphere:
\be \label{unitcond}
\ab^{\prime 2}=\bb^{\prime 2}= 1\;.
\ee
Notice that while the periodicity of $\ab$ and $\bb$ is $L$, the
periodicity of the actual motion of the string is $L/2$, since
$\r(\sigma+L/2,\tau+L/2)=\r(\sigma,\tau)$.
There is also an additional constraint that $\langle\ab'\rangle
=\langle\bb'\rangle=0$, coming from the facts that the loop
is closed and that we are in the {\it c.o.m.}\ frame.
Thus the string motion is specified by two curves on the unit
sphere which must have zero weight. 

Kinks, unlike cusps, which occur
whenever these curves intersect, form when one of $\mathbf{a}'$, 
$\mathbf{b}'$ or one of their derivatives has a discontinuity. 
As well as being ubiquitous on cosmic strings, kinks can also prevent the
formation of cusps on cosmic string loops, as a discontinuity on either
$\mathbf{a}'$ or $\mathbf{b}'$ allows the curves to miss each other if 
they would otherwise have intersected at that point.
Kinks, like cusps, emit bursts of gravitational radiation whose signal was
calculated initially by Vachaspati and others, \cite{FAN,GV}, and more
rigorously by DV, \cite{DV}, as we now briefly review.

In order to obtain the gravitational waveform of the kink GWB, 
we begin with the linearised Einstein equations 
\begin{equation}
\Box {\bar h} _{\mu\nu} = -16\pi G T_{\mu\nu}
\end{equation}
where ${\bar h}_{\mu\nu}$ is the trace reversed metric perturbation,
and $T_{\mu\nu}$ is the string energy momentum
\begin{equation}
T^{\mu\nu} ({\bf k},\omega) = \frac{\mu}{L} \int_0^L d\sigma_+
\int_0^L d\sigma_- {\dot X}^{(\mu}_+ {\dot X}^{\nu)}_-
e^{-\frac{i}{2}(k\cdot X_+ + k\cdot X_-)}
\end{equation}
written in momentum space with $k^\mu = \frac{4\pi m}{L} (1, {\bf n}) 
=m\omega_L (1,{\bf n})$, where $\omega_L$ is the frequency of the 
fundamental mode of the string loop, and $X_\pm$ denote the position vectors
$X^\mu_\pm = \left (\sigma_\pm, \ab / \bb (\sigma_\pm) \right)$
associated with the left and right moving modes ($X^\mu = (X_+^\mu+X_-^\mu)/2$.
In the far field approximation the solution is given by
\begin{equation}\label{waveform}
{\bar h} _{\mu\nu} \simeq \frac{4G}{r} \sum_\omega
e^{-i\omega(t-r)} T_{\mu\nu} ({\bf k}, \omega) \;,
\end{equation}
thus finding the perturbation reduces to determining the integrals
\begin{equation}\label{Iintegrals}
I_\pm^\mu = \int_0^L d\sigma_\pm {\dot X}^{\mu}_\pm e^{-\frac{i}{2}k\cdot X_\pm}
\end{equation}
which appear in the energy momentum tensor.

The key first step of the DV calculation is to determine when these
integrals are not exponentially suppressed, and to remove the leading
order gauge dependence.  Without loss of generality in what follows,
we will assume that the discontinuity is in the $\ab'$ curve, and that
the $\bb'$ curve is continuous.

First, note that around $\sigma_+$, we can expand the right moving 
position vector $X^\mu_+$ as
\be
X^\mu_+ (\sigma_+) = \ell^\mu \sigma_+ + \frac{1}{2}
{\ddot X}^\mu _{_0+} \sigma_+^2 +
\frac{1}{6} {\ddot X}\hskip -2mm{\dot{\phantom{O}}}\!^\mu_{_0+}
\sigma_+^3 \label{DVexpX}
\ee
where the subscript $0$ refers to evaluation at $\sigma_+=0$.
Defining the angle between $k^\mu$ and $\ell^\mu$ as $\theta$, which
is assumed to be small, and writing $d^\mu = k^\mu-\ell^\mu = (0,{\bf d})$
(where $|{\bf d}| \simeq \theta$), and using the gauge conditions, DV
obtain:
\be
k_\mu X_+^\mu = m\omega_L \left [ \frac{1}{2} \theta^2 \, \sigma_+
- \frac{1}{2}{\bf n} \cdot {\bb}'' \, \sigma_+^2
+ \frac{1}{6} \left ( {\bb}^{\prime\prime2} - {\bf d} \cdot {\bb}'''
\right ) \sigma_+^3 \right ]
\label{DVkdotX}
\ee
(although they drop the ${\bf d}\cdot \bb'''$ term as it is subdominant).
Thus the integral takes the form
\be
I^\mu_+  = \int [ k^\mu - d^\mu + {\ddot X}^\mu_+ \sigma_+ ]
\exp \left [ - \frac{im\omega_L}{12}
( 3 \theta^2 \sigma_+ - 3 \theta |{\ddot X}_+| \sigma_+ ^2
\cos\beta + |{\ddot X}_+|^2 \sigma^3_+ ) \right] d\sigma_+
\label{DVI}
\ee
where $\beta$ is the angle between $\bf d$ and $\bb''$.
As Damour and Vilenkin pointed out, the first $k^\mu$ term is a pure
gauge, (although there are subtleties when $d^\mu \neq 0$) and therefore
the physical part of the integral is proportional to ${\ddot X}_+^\mu$.

For $\theta=\beta=0$, this integral 
can be evaluated analytically (see also \cite{CCGGZ}) as
\be
I^\mu_+ = - \left ( \frac{12}{m\omega_L {\ddot X}^2_+} \right ) ^{2/3}
\frac{i}{\sqrt{3}}\; \Gamma \left ( \frac{2}{3} \right ) {\ddot X}^\mu_+.
\ee
Note that the origin of $\sigma_+$ is arbitrary, hence
``$\theta = 0$'' means that the wave vector coincides with the 
velocity of the right mover for some value of $\sigma_+$, indeed
a shift in origin of $\sigma_+$ simply corresponds to introducing
a disparity $d^\mu = - \delta\sigma_+ {\ddot X}_{_0+}$, thus
the ``$\beta=0$" saddle is simply reflecting this gauge invariance.

We therefore arrive at the conclusion that the contribution to the 
gravitational perturbation for the (continuous) right moving mode 
has a power law decay of $f^{-2/3}$, and
is strongly localised around the velocity vector for that mode.
As shown in DV the angular spread transverse to the $\bb'$ trajectory
is 
\be
\theta_m =  \left ( \frac{12{\ddot X}}{m\omega_L} \right ) ^{1/3}
\simeq \left (  \frac{2}{Lf} \right ) ^{1/3}.
\label{thetamaxdef}
\ee
Thus the integrals $I_\pm$ are enhanced for wave vectors coincident with
the wave on the string. Clearly, when these left and right moving
vectors coincide, this corresponds to both $I_+$ and $I_-$ being
on their saddle points which gives a strong signal sharply localised
in a solid angle $\pi \theta_m^2$ around this direction. However,
since we are considering a kink, we must assume that the discontinuity
causes the $\ab'$ curve to `miss' the $\bb'$ curve, and thus we must
look more closely at $I_-$, which requires a slightly
different approach. 

To compute the kink integral, DV split it into three parts around
the discontinuity at $\sigma_-=\sigma_-^{disc}$, taking 
$\dot{X}_-^\mu$ to jump from $n_1^\mu = (1, {\bf n}_1)$ 
at $\sigma_-^{disc}-0$ to $n_2^\mu = (1, {\bf n}_2)$, 
at $\sigma_-^{disc}+0$ where ${\bf n}_{1/2}$ are unit 
vectors.
The integral becomes:
\begin{align}\label{Iminus}
I_-^\mu &= \int_{\sigma_0}^{\sigma_-^{disc}-\epsilon} d\sigma_- {\dot
X}^{\mu}_-e^{-\frac{i}{2}(k\cdot X_+ + k\cdot X_-)}
+\int_{\sigma_-^{disc}-\epsilon}^{\sigma_-^{disc}+\epsilon} d\sigma_- {\dot
X}^{\mu}_-e^{-\frac{i}{2}(k\cdot X_+ + k\cdot X_-)} \nonumber\\
&+\int_{\sigma_-^{disc}+\epsilon}^{\sigma_0+L} d\sigma_- {\dot
X}^{\mu}_-e^{-\frac{i}{2}(k\cdot X_+ + k\cdot X_-)}\;,
\end{align}
where $\epsilon$ is an arbitrary small parameter, introduced to separate out the
discontinuity from the rest of the integral (i.e.\ the middle term of
(\ref{Iminus})), which will be allowed to go to zero at the end of the
calculation. The first and third terms then cover the remainder of the
integration range on either side of the discontinuity.

The terms in (\ref{Iminus}) which no longer contain the
discontinuity only have a significant contribution when the
wavevector is parallel to the velocity vector, however, by
assumption, this will not correspond to a saddle point of the $I_+$
integral, otherwise we would have a double saddle and hence a cusp.
Thus we must take $k^\mu$ to lie outside the range
$\pm \theta_m$ of the wave vector ${\dot X}_-^\mu$, and therefore
the first and third terms in eqn.\;(\ref{Iminus}) can be neglected.
Eqn.\;(\ref{Iminus}) can therefore be evaluated to leading order as
\bea
I_-^\mu &\simeq& \int_{\sigma_-^{disc}-\epsilon}^{\sigma_-^{disc}}n_1^\mu
e^{-\frac{im\omega_L}{2}\hat{k}\cdot n_1\sigma_-} -
\int_{\sigma_-^{disc}+\epsilon}^{\sigma_-^{disc}}n_2^\mu
e^{-\frac{im\omega_L}{2}\hat{k}\cdot n_2\sigma_-} \\
&=& -\frac{2n_1^\mu}{im\omega_L k\cdot n_1} \;
e^{-\frac{im\omega_L}{2}k\cdot n_1\sigma_-}\bigg\rvert
_{\sigma_-^{disc}-\epsilon}^{\sigma_-^{disc}} -
\bigg(-\frac{2n_2^\mu}{im\omega_L \hat{k}\cdot n_2}\bigg)
e^{-\frac{im\omega_L}{2}\hat{k}\cdot n_2\sigma_-}\bigg\rvert
_{\sigma_-^{disc}+\epsilon}^{\sigma_-^{disc}}\; \nonumber \\
&\simeq& \frac{2i}{m\omega_L}\bigg(\frac{n_1^\mu}{\hat{k}\cdot
n_1}-\frac{n_2^\mu}{\hat{k}\cdot n_2}\bigg)\;
= \frac{2i}{m\omega_L} v^\mu \label{IminusSoln}
\eea
where $\hat{k}=(1,\bf{n})$, so that $m\omega_L\hat{k}^\mu = k^\mu$.

Putting together the $I_\pm$ integrals, DV argued that the 
logarithmic kink waveform, 
$h^{\mu\nu}(f) = 2G\mu|f|I_+^{(\mu}I_-^{\nu)}/r$,
is therefore
\begin{equation}\label{kinkwform}
 h^{kink}(f,\theta) \sim \frac{G\mu 
L^{1/3}}{r|f|^{2/3}} H[\theta_m - \theta]\;.
\end{equation}

To transform (\ref{kinkwform}) to a cosmological setting,
we replace $f$ by $(1+z)f$, where $z$ is now the redshift at 
the time of emission of the kink burst, and $r$ by the
physical distance 
\be
a_0 r = a_0 \int_{t_e}^{t_0} \frac{dt}{a} = \int_0^z \frac{dz}{H}
= (1+z) D_A(z)
\ee
where $D_A(z)$ is the angular diameter distance at redshift $z$.

To estimate the GWB signal rate from the cosmological 
network, DV assumed the one scale model, in which the network is
approximated by the dominant scale set by the cosmological time:
\be
L \sim \alpha t \qquad , \qquad \; n_L(t) \sim 1/ (\alpha t^3)
\label{onescale}
\ee
for the length and number density of the string network at
cosmological time $t$. Here $\alpha \sim \Gamma G\mu$ is a numerically
determined constant, \cite{gwave,GWL,Eloss, recentalpha}, presumed to 
represent the rate of energy loss from string loops via gravitational 
radiation, where, as in DV, we take $\Gamma \sim 50$. 

Finally, the rate of GWB's from kinks observed in the spacetime volume in
redshift interval $dz$ around a frequency $f$  can be calculated from
\begin{equation}\label{dndotkink}
d\dot{N}_k = \frac{2\nu_k(z)}{1+z}\frac{\pi\theta_mD_A(z)^2}{(1+z)H(z)}dz\;,
\end{equation}
where the number of kink events on a loop $\nu_k$ is given by
\be
\nu \sim \,{\cal K}\, \frac{n_L}{P T_L} \sim \frac{2{\cal K}}{P\alpha^2 t^4}
\label{nudef}
\ee
with ${\cal K}$ being the average number of kinks per loop. 

In order to perform this calculation, DV made several simplifications
concerning the kink GWB which we will now critically examine. First of all,
in deriving the kink GWB amplitude (\ref{kinkwform}) the amplitude
of the integral $I_-$ was taken to be ${\cal O}(1)$. 

In order to get a feel for the kink signal, consider the magnitude of
$v^\mu$ in (\ref{IminusSoln})\footnote{Recall that we have already
used gauge invariance to remove the $k^\mu$ factor in $I_+$.}. 
Choosing coordinates for the Kibble-Turok 2-sphere so that 
the relevant kink and gravitational wave 3-vectors are:
\bea
{\bf n}_1 &=& \cos \frac{\delta}{2} \; {\bf e}_1 
+ \sin\frac{\delta}{2} \; {\bf e}_2 \nonumber \\
{\bf n}_2 &=& \cos \frac{\delta}{2} \; {\bf e}_1 
- \sin\frac{\delta}{2} \; {\bf e}_2 \\
{\bf n} &=& \sin\theta \cos\phi\; {\bf e}_1 + \sin\theta\sin\phi\; {\bf e}_2
+\cos\theta \; {\bf e}_3 \nonumber
\eea
where $\delta$ is the {\it discontinuity angle}, i.e.\ the
angle between ${\bf n}_1$ and ${\bf n}_2$. The `amplitude' 
of the kink can therefore be straightforwardly calculated as:
\be
{\cal A} = |v^\mu v_\mu|^{1/2} = \frac{2 \sin\delta/2}
{[1-\sin\theta\cos(\phi+\delta/2)]^{1/2}
[1-\sin\theta\cos(\phi-\delta/2)]^{1/2}}.
\label{kinkamp}
\ee
Note that this differs from the definition of the amplitude in \cite{BBHS,ED}
in two ways. First of all, we have taken the magnitude of the 4-vector $v^\mu$,
this is important because the worldsheet is a relativistic object, and the
gravitational radiation is determined by the energy momentum tensor. The
4-vector component is in fact related to terms appearing in the extrinsic
curvature of the worldsheet at the kink, (see appendix of \cite{CCGGZ}), 
and is thus
the appropriate relativistic quantity to consider. In addition, \cite{BBHS}
explicitly removed the denominator from consideration, arguing that it
was ${\cal O}(1)$. While this is generally true, it is equally true of the
numerator in $v^\mu$, and both contribute to the overall amplitude.
We therefore arrive at the result that the amplitude varies by more than 
an order of magnitude over the sphere, and, depending on the $\theta_m$ 
cutoff, potentially significantly more.

By plotting the density of ${\cal A}$ on the unit sphere and combining with 
the saddle of the $\bb'$ curve, the gravitational radiation from the kink
can be seen to be spread out along the $\bb'$ curve, but does peak
significantly in the region where that curve runs between the two points 
marking the kink discontinuity. By supposition, 
the curve must run between the discontinuity (otherwise a cusp would result) 
as it must have zero weight, $\langle \bb' \rangle = {\bf0}$.

In figure \ref{fig:kinkamp} the amplitude is shown along a great
circle running between the discontinuity for a range of discontinuity
angles. We have chosen a generic
great circle which peaks at a latitude of $30^o$, and crosses
the equator at a longitude of $18^oE$. The figure shows both the large
range in the amplitude ${\cal A}$, as well as the focussing along
the curve. For example, the black curve, corresponding to a discontinuity
angle of $\delta = \pi/3$, is sharply peaked between $\lambda \sim 3\pi/8
\to5\pi/8$ corresponding to a genuine fan of radiation. The figure
also shows clearly that as the discontinuity angle increases, ${\cal A}$ 
decreases between the discontinuity points, and thus contrary to \cite{BBHS}, 
we find that the {\it closer} these points are, the more strongly 
focussed the gravitational radiation and the more power it has. 
\FIGURE{
\label{fig:kinkamp}
\includegraphics[height=6cm]{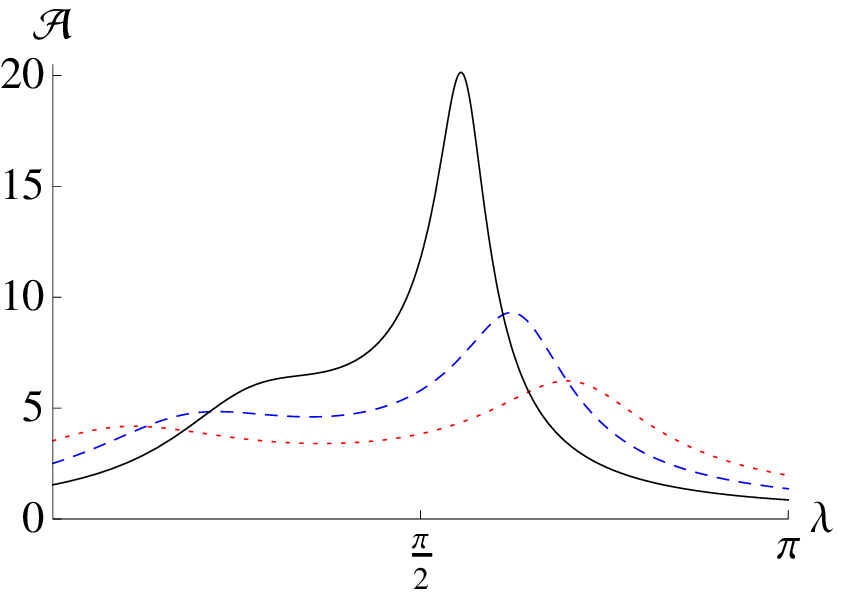}
\caption{The kink amplitude (\ref{kinkamp}) along a great circle
for various angles of separation for the discontinuity as a function
of the proper distance along the path. In each case the great circle
runs from $\theta = \pi/3$ to $2\pi/3$, and crosses the equator at
$\phi = \pi/10$. The dotted red line corresponds to a discontinuity angle 
of $\delta = 2\pi/3$, the dashed blue line to $\delta =\pi/2$, and the solid
black line to $\delta =\pi/3$.
} }

The second simplification DV make is in the solid angle integration of the 
GWB rate (\ref{dndotkink}). Here, the integral of the kink GWB is assumed
to cover a solid angle $2\pi\theta_m$. This reflects the integration over
the fan, which in this instance is approximated by a great circle (or the
fundamental mode). Since the kink signal peaks for wavevectors
in the region between the discontinuity, the actual solid angle of the 
fan is less than $2\pi\theta_m$.

In order to confirm that these effects do not significantly change the 
DV result, we took various weighted averages of the kink signal. First,
the most straightforward, we simply integrated the amplitude over the
sphere to find an average amplitude per unit solid angle. This varies
remarkably little as the discontinuity angle varies.
However, as the $\bb'$ curve must run between the ${\bf n}_i$,
a more accurate measure might be to take into account the localisation 
of the GWB on the fan, and average over all possible ``fans'' running
between the discontinuity. Using
great circle trajectories such as those in figure \ref{fig:kinkamp},
the result is again a mild dependence on the discontinuity angle, but an
anti-correlation in this case. The clear outcome is that
the weighting of the integral due to the amplitude and geometry of a
kink is indeed well approximated by a number of ${\cal O} (1)$ in
a solid angle of ${\cal O}(\pi \theta_m)$.

To recap: the DV computation shows 
due to the mobile and persistent nature of a kink,
the emission is not beamed in a cone like the cusp, 
but is instead radiated in a one dimensional `fan-like' set of 
directions, \cite{DV}, which arise from the occurrence of saddle points;
the inclusion of the angle $\theta_m$, so that the observation direction is
slightly offset from the direction of the emission, results in the 1D `line' of
emission from the kink being widened so that a strip is swept out on the surface
of the unit sphere and the volume factor becomes $\theta_m/2$, \cite{OMS}.

\section{Kinks in Higher Dimensions}

The effect of extra dimensions on the formation of, and gravitational 
radiation from, cusps was explored in \cite{CCGGZ}, where it was 
found that one of the main differences in looking at cusps with extra 
dimensions was the fundamental change in behaviour of 2D curves
moving on the surface of a $2+n$-sphere rather than a 2-sphere, 
resulting in intersections, which had been generic, becoming extremely rare. 
A kink, however, does not require the curves to intersect, nor will the 
appearance of the discontinuity depend on the number of dimensions in 
which the (super)string moves; kink
formation will thus be unaffected by the inclusion of extra dimensions. 

This is not to say that the presence of extra dimensions will have no effect on
the GWB's emitted by kinks. The change in the intercommutation probability 
due to the presence of extra dimensions, \cite{ICSC}, affects the 
network density, leading to an enhancement of the kink amplitude.
Furthermore, the slowing down of the wave velocity in our noncompact
dimensions affects the computation of the saddle point integral, $I_+$, 
which will affect the kink GWB as it does the cusp,
where it contributed to a reduction in the radiation cone, \cite{CCGGZ}.

Following the notation of \cite{CCGGZ}, we consider (super)string 
solutions in $4+n$ dimensions, which can be expressed in 
Kibble-Turok notation as 
\be
{\bf R}= \frac{1}{2}[\Ab(\sigma_-)+\Bb(\sigma_+)],
\ee
where the upper case denotes the full ($3+n$) dimensional spatial 
vectors and $|\Ab'|^2 = |\Bb'|^2 = 1$ from the gauge choice as before. 
Since a kink simply requires a discontinuity in one of the left- or 
right- moving curves, the fact that the three dimensional part of
these (still written as $\ab',\bb'$) no longer lies on a unit $S^2$
has no relevance for kink formation in the extra dimensions. 
The computation of the $I_-$ integral, dominated by the endpoints of
the discontinuity, will also be unaltered.

However, the saddle point integral $I_+$ will damp too quickly unless we
take the extra dimensional component, ${\mbox{\textit{b}}}\ll 1$, where
$|\bb'|^2 = 1-b^2$. Combining this with the estimation
of ${\bf n}' . \bb'' = {\cal O} ({\mbox{\textit{b}}}) |{\ddot X}|$ 
following the methods of \cite{CCGGZ}, the exponent of $I_+$ is:
\begin{equation} 
k_\mu X_+^\mu = \frac{1}{2} ( \theta^2 + {\mbox{\textit{b}}}^2) \, \sigma_+ -
\frac{1}{2}\left ( \theta + {\mbox{\textit{b}}} \right ) |{\ddot X_+}| \,
\sigma_+^2
+ \frac{1}{6} {\ddot X_+ }^2 \, \sigma_+^3\;,
\end{equation}
thus the angle $\theta_m$  narrows as a result of the
inclusion of extra dimensions, which will affect the kink GWB signal. 

We therefore obtain that, as for the cusp \cite{CCGGZ}, the 
extra dimensional (logarithmic) kink waveform is the same as 
the 3D waveform, eqn.\;(\ref{kinkwform}),  with
a reduced beaming angle:
\begin{equation}\label{EDkinkwform}
 h^{kink}(f,\theta) \sim \frac{G\mu L^{1/3}}{r|f|^{2/3}}
H[\theta_{{\mbox{\textit{b}}}} - \theta]\;,
\end{equation}
where
 \begin{equation}
\theta_{{\mbox{\textit{b}}}} = \theta_m - {\mbox{\textit{b}}} \simeq \left ( 
\frac{2}{Lf} \right ) ^{1/3} -  {\mbox{\textit{b}}}\;.
\label{thetagendefKink}
\end{equation}

We now need to compute the event rate of these higher dimensional kink
GWB's for the cosmological (super)string network. Since the mechanism 
which produces kinks is unchanged by the inclusion of extra dimensions, 
we assume that the average number of kinks in a loop, $\mathcal{K}$, 
is unchanged in the higher dimensional setting and will not depend on
the number of extra dimensions, (unlike cusps as discussed in \cite{CCGGZ}). 
However, the (super)string network will possess kinks with a range of 
different $\theta_{\textit{b}}$ values; we therefore write 
\begin{equation}
\frac{d{\dot N}_k}{dz\, d\textit{b}} \sim \frac{2{\cal K} n_L(z)}{P
T_L(z)}\frac{\pi \left (\theta_m(z) - \textit{b}\right ) D_A(z)^2}{(1+z)^2
H(z)}\;.
\label{NdeltaKink}
\end{equation}
Integrating over $\textit{b}$ thus yields the kink GWB rate 
\begin{equation}\label{EDkinkRate}
\frac{d{\dot N}_k}{dz} \sim \frac{{\cal K}\theta_m^2 n_L(z)}{P T_L(z)}\frac{\pi
D_A(z)^2}{(1+z)^2 H(z)}\;.
\end{equation}
\FIGURE{\label{Fig:KinkAmp}
\includegraphics[width=7cm]{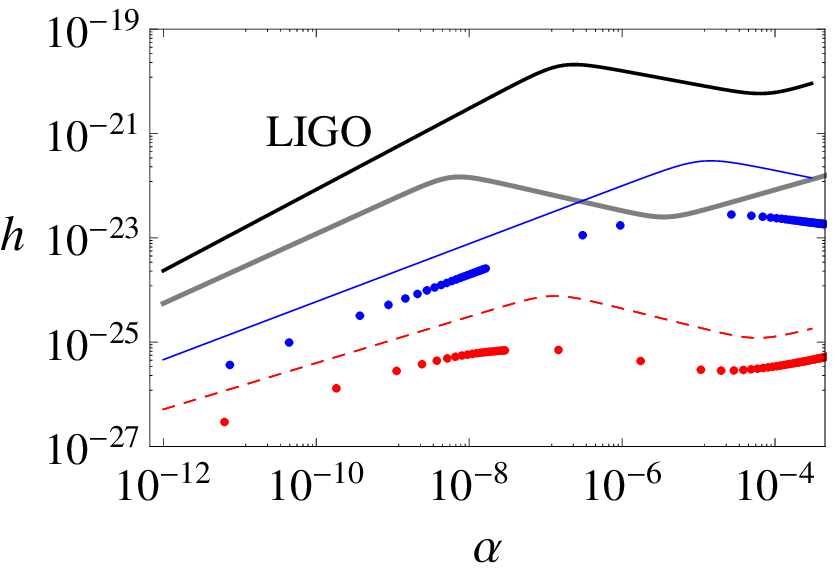}\nobreak
\includegraphics[width=7cm]{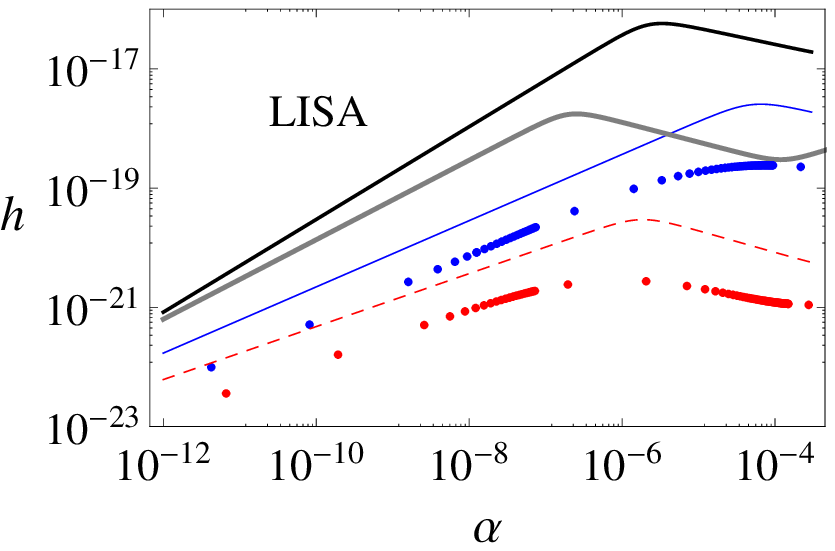}
\caption[Kink gravitational wave burst only (LIGO)]{Log-log plots of
the GWB amplitude as a function of $\alpha$ for the LIGO ($f=150$ Hz)
and LISA ($f = 3.9$mHz) frequencies at
a detection rate of 1 per year. The original DV computation is
shown as a solid blue line, and the kink amplitude with extra dimensions
is shown as a dashed red line. The sets of individual dots
correspond to the exact numerical redshift integrations.
For comparison, the cusp amplitude computed using interpolating
functions is also shown in solid thick black for no extra dimensions,
and thick grey for one additional extra dimension. All plots use an
intercommutation probability of $P=10^{-3}$.}
}

Figure \ref{Fig:KinkAmp} shows the gravitational wave amplitude 
for the cosmic string kink burst calculated by DV, \cite{DV}, along 
with our higher dimensional results for the characteristic
frequencies of the LIGO and LISA gravitational wave detectors. 
We present the amplitudes calculated using the DV interpolating
functions (neglecting $\Omega_\Lambda$), and also include the 
results of an exact numerical integration performed for the
concordance cosmology ($\Omega_r = 4.6 \times 10^{-5},\, \Omega_m = 0.28,\,
\Omega_\Lambda =1-\Omega_m-\Omega_r$). For comparison, we also 
include the equivalent plots for the cusp, with the amplitudes for
zero and one extra dimension shown. A fixed value, $P=10^{-3}$ was
chosen for all the plots. Note that there is
no dependence on the number of extra dimensions with the kink amplitude,
but for cusps, with each additional extra dimension, the amplitude is 
suppressed by approximately one order of magnitude. 

\FIGURE{\label{Fig:RateKink}
\includegraphics[width=7cm]{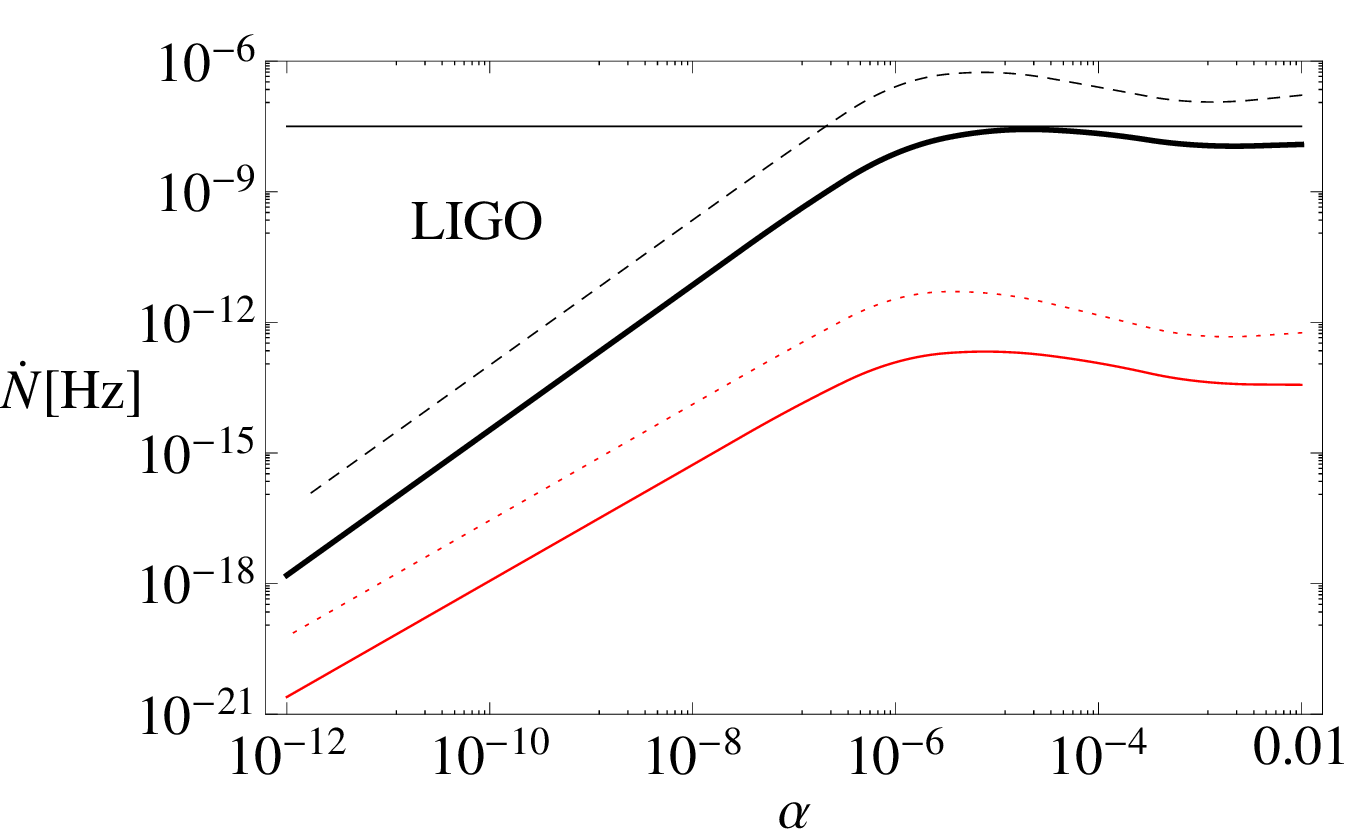}
\includegraphics[width=7cm]{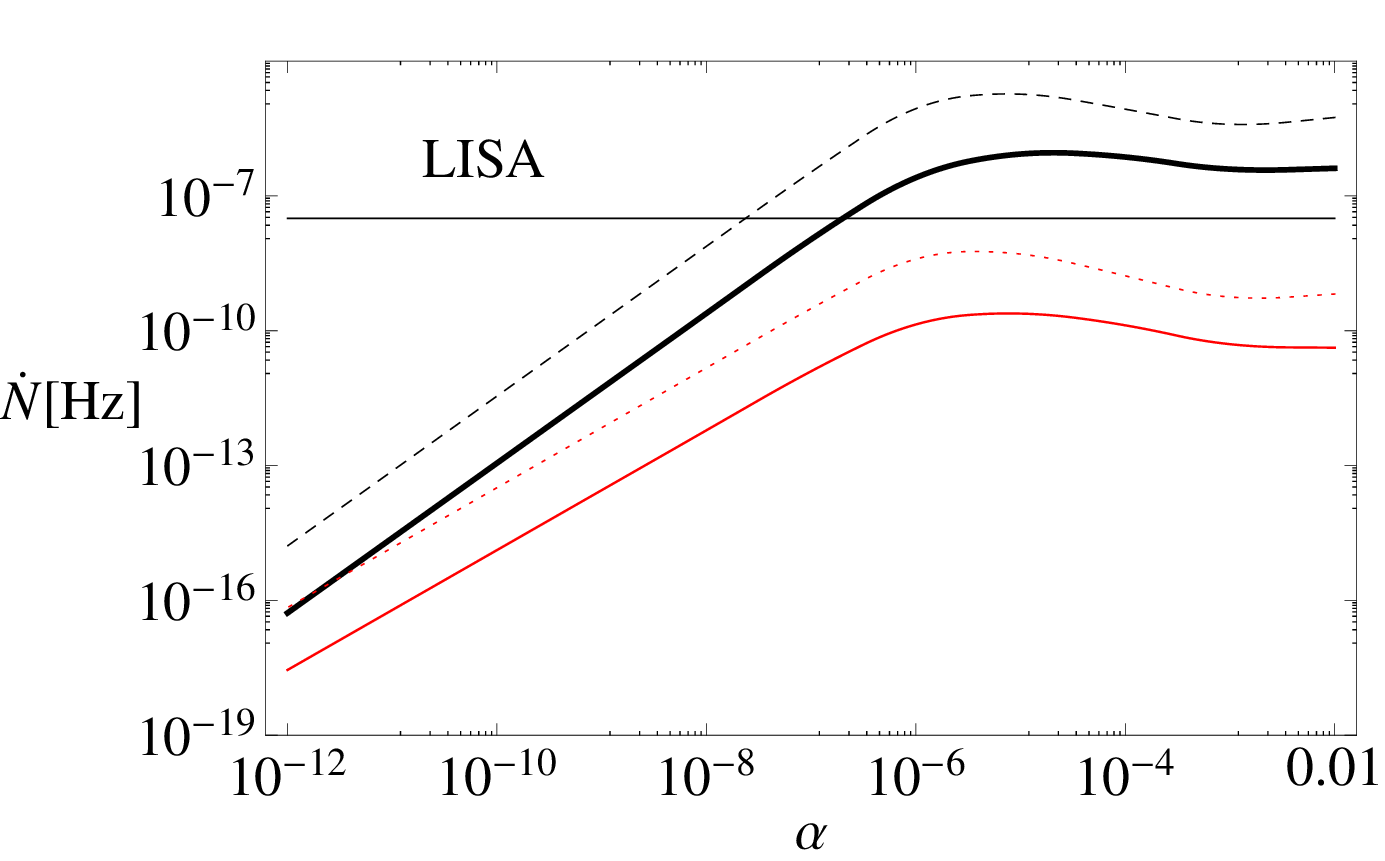}
\caption[Rate of gravitational wave burst  from kinks only (LIGO)]{Similar
plots to those of figure \ref{Fig:KinkAmp} but in this case showing the
expected rate at an amplitude of $10^{-21}$ s$^{-1/3}$, using the method of
\cite{Siem}. In this case, the thick
solid lines are the numerical integration results. From top to bottom: the
DV kink, (black,solid) and the extra dimensional kink, (red, dashed). Both have
an intercommutation probability of $P=10^{-3}$. The horizontal black line
indicates a rate of one event per year.}
}

An alternative useful way of displaying the kink GWB is to instead plot
the expected rates of detection at a given amplitude as described in 
\cite{Siem}. The one scale model is something of an oversimplification
(though the only sensible analytic approximation to date), and in \cite{Siem},
the authors show how to have full dynamical range for the network, including
a dependence in $h$ on the length distribution in the network, giving a
rate per unit redshift, per unit amplitude. Integrating out using the one
scale model, which essentially associates redshift with amplitude, gives
an expected detection rate at a particular amplitude. Figure
\ref{Fig:RateKink} shows this rate calculation for the kinks with extra
dimensions in the frequency range of LIGO and LISA. In
this case, the amplitude of the kink waveform is found by comparison with
\begin{equation}
h^{kink}=A |f|^{-2/3}\;,
\end{equation}
so that the redshift values used for the integration at various values of $G\mu$
are found from 
\begin{equation}
\frac{\varphi_t^{1/3}(z)}{(1+z)^{2/3}\varphi_r(z)}
=\frac{50A H_0^{-2/3}}{\alpha^{4/3}}\;,
\end{equation}
where we once again use $A=10^{-21}$ s$^{-1/3}$,  $\alpha\sim 
50 G\mu$ and $\varphi_t$, $\varphi_r$ are either the
DV interpolating functions or related to the exact functions $t(z)$, $D_A(z)$, 
(c.f. Siemens et al., \cite{Siem}).

\section{Discussion}

We have computed the gravitational wave signal due to cosmic
(super)string kinks, and have found that there is a dimension
independent suppression of approximately an order of magnitude 
compared to the purely four-dimensional result. The fact that
the suppression is independent of the number of internal dimensions
is in contrast to the cusp signal, \cite{CCGGZ}, which drops
sharply with each extra internal dimension. The reason for this
difference is that the cusps have an additional probabilistic
suppression due to the factors determining their formation.

Motion in the extra dimensions slows down the (super)string in the 
non-compact dimensions, and for the cusp, which requires that
the (super)string reach the speed of light due to a constructive 
interference effect, this has a strong impact on the likelihood
of formation. Kinks in contrast require only a discontinuity 
in the wave-vectors of the (super)string, and will persist even if
those wave-vectors are not quite null. Moreover the mechanisms
for kink formation (self-intersection, intercommutation) are
not qualitatively affected by extra dimensions, thus the
(super)string network will still have many kinks.
Although the kink signal is at
least an order of magnitude lower than the cusp signal 
before the extra dimensions are taken into account,
the ubiquitous nature of kinks on a (super)string loop in extra
dimensions, as compared to the near cusp events 
discussed in \cite{CCGGZ}, makes the kink amplitude an
important signal, particularly if the number of extra dimensions is large.
A particularly important point to note is that, like DV, we have assumed 
a value of ${\cal K} \simeq 1$. Clearly this is a gross underestimate
of the true gravitational signal, the signal rises approximately one
order of magnitude for every three orders of magnitude increase in ${\cal K}$.
If kinks indeed proliferate on (super)strings, the real picture may in
fact be extremely optimistic for (super)string detection.

A by-product of our analysis was that in computing the kink
amplitude in section \ref{sec:3d}, we carefully followed the
magnitude of the discontinuous wave integral $I_-$. Previously
in the literature, \cite{ED,BBHS}, the amplitude had been defined
by taking the 3-vector and ignoring a dividing factor. By using
the full relativistic expression, and keeping all factors, we
showed that the actual picture for the kink amplitude was rather
different. In contrast to \cite{BBHS}, we find that very `sharp'
(where sharp refers to the three dimensional profile and corresponds
to a large discontinuity angle)
kinks do not radiate more than apparently milder kinks.

It is worth explaining why this is so, as it can seem counter-intuitive.
The real key to gravitational radiation is the relativistic
nature of the source. Large discontinuity kinks move slowly
along the (super)string, and thus although they are sharp in the three
dimensional sense, they do not move fast. In fact, the less `sharp' 
kinks move much faster, hence radiate more. Of course, this modelling
is looking at an individual kink, in reality, the loops will have many 
kinks, and there may also be many sharp kinks moving rapidly. However,
we believe this is best reflected in an increase in the overall 
GWB amplitude via ${\cal K}$, rather than in the individual kink
burst. 

We have also not explored the effect of junctions in detail. Cosmic
(super)strings have the additional possibility of having junctions due
to the microphysics inherited from the underlying type IIB string
theory \cite{PQ}. These junctions can effect the network
evolution, \cite{AC}, as well as having gravitational radiation
of their own \cite{GWCSS}. Most recently, it has been shown that
the kinematics of the junctions \cite{JCTkin} can lead to a proliferation
of kinks, \cite{BBHS}, as a kink hitting a junction both reflects and
transmits through the junction. All of these effects indicate that
kinks are possibly a far more important feature of (super)string
networks that the ordinary type of cosmic string.

It is worth noting that the effects of extra dimensions are 
dependent on the frequency at which they are observed; it is evident, 
(particularly from the rate plots in figure \ref{Fig:RateKink}), that there
is a much better chance of observing a kink signal at LISA rather than LIGO,
although the approximation used in this method
to find the waveform, (i.e.\ that $\theta_m \ll 1$), breaks down at nHz
frequencies, where the angle $\theta_m \sim \mathcal{O}(1)$ and the integral
$I_+\to 0$ rapidly. At lower frequencies, such as those probed
by the pulsars for example, \cite{pulsar}, the effect of the extra dimensions 
essentially washes out, and at this level, the (super)string
kinks will simply contribute to the general
stochastic GW background, \cite{OMS}.

To sum up, although we find a significant suppression of the GWB
signal from kinks due to extra dimensions, the overall suppression
is much less than that for cusps. Thus, although the kink signal 
is weaker than that of the cusp, its independence of the number 
of extra dimensions, $n$, could lead to it being important, 
particularly if $n$ is large. If, in addition, there are other
kinematical effects driving kink formation, then the outlook for
detection at LIGO becomes extremely optimistic over a wide range
of (super)string tensions (although the current bounds, \cite{obsv}, do 
need to be revisited!). Clearly it is well worth a more detailed and
comprehensive modelling of kinks on networks to give a definitive
computation of the gravitational wave signal. 

\acknowledgments

We would like to thank Sarah Chadburn, Ghazal Geshnizjani, Dani
Steer, and Ivonne Zavala for useful conversations.
EOC is supported by the EC FP6 program through the 
Marie Curie EST project MEST-CT-2005-02174, and RG
by STFC under rolling grant ST/G000433/1.

\end{document}